\documentclass[aps,pre,floats,showpacs]{revtex4-1}
\usepackage{epsfig}
\usepackage{amsmath}
\usepackage{amssymb}
\usepackage{amsbsy}
\usepackage{graphicx}
\usepackage{color}
\newcommand{\ve}{\varepsilon}

\newcommand{\vp}{\varphi}

\newcommand{\mm}[1]{\mathrm{#1}}

\newcommand{\sech}{\mm{ \, sech}}
\begin{document}
\title{Soliton Propagation through a Disordered System:\\
Statistics of the Transmission Delay}
\author{Sergey A. Gredeskul$^{1,2}$, Stanislav A. Derevyanko$^{3}$,
 Alexander S. Kovalev$^4$, Jaroslaw E. Prilepsky$^{3,4}$}

\affiliation{$^1$Department of Physics, Ben Gurion University of
Negev, Beer Sheva, Israel\\
$^2$ Nonlinear Physics Center, Australian National University,
Canberra, Australia\\
$^3$ Nonlinearity and Complexity Research Group, Aston University,
Birmingham, UK\\
$^4$ B.I. Verkin Institute for Low Temperature Physics and
Engineering, NASU, Kharkov, Ukraine}


\begin{abstract}
We have studied the soliton propagation through a segment containing random point-like scatterers. In the limit of small concentration of scatterers when the mean distance between the scatterers is larger than the soliton width a method
has been developed for obtaining the statistical characteristics of the soliton transmission through the segment. The method is applicable for any classical particle traversing through a disordered segment with the given velocity transformation after each act of scattering. In the case of weak scattering and relatively short disordered segment the transmission time delay of a fast soliton is mostly determined by the shifts of the soliton center after each act of scattering. For sufficiently long segments the main contribution to the delay is due to the shifts of the amplitude and velocity  of a fast soliton after each scatterer. Corresponding crossover lengths for both cases of light and heavy solitons have been obtained. We have also calculated
the exact probability density function of the soliton transmission time delay for a sufficiently long segment. In the case of weak identical scatterers the latter is a universal function which depends on a sole parameter -- mean number of scatterers in a segment.
\end{abstract}
\maketitle
\section{Introduction}\label{sec:intro} \thispagestyle{headings}
The interplay between disorder and nonlinearity has attracted the attention of
physicists for more than twenty years \cite{NL&D,KG92}. The latest advancements
in experimental techniques and computational facilities have brought about
a new peak of interest in the problem and already led to a wider
understanding of the corresponding  phenomena \cite{HaifaWorkshop}.
Physical properties of disordered nonlinear systems reveal much
more sensitivity to the details of the formulation of
the problem than their linear counterparts. For example, in nonlinear systems there exist
three different transmission regimes: beside the
exponential decrease of the transmission coefficient with the
length of the disordered segment \cite{KGSV,Hopkins} (which is the sole
possibility in the linear case), power law decrease can occur \cite{Devillard}, and, in the case of sufficiently strong
nonlinearity, the transmission coefficient may not even decrease at
all \cite{KGSV,Hopkins}.

One of the main concepts in the theory of nonlinear waves is a concept of a soliton \cite{zs71,mnp84,km89} -- a particle-like stable nonlinear excitation observed in many nonlinear physical systems. A particular example revealing the reciprocal action between the nonlinearity and disorder is a soliton propagation in random media. One can observe here an entire plethora of problems which differ from each other by (i) the type of nonlinearity (e.g. nonlinear Schr\"{o}dinger equation \cite{kkc87,KGSV,Burtsev,Knapp,BronskiAn,BronskiNum,GarnierSIAM,GarnierWRM}, sine-Gordon equation \cite{KFV,FKV0,Gonz03}, so-called $\phi^4$-system \cite{KFV,FKV1,Kalbermann}), (ii) the way in which randomness enters the system (e.g. in the random potential form  \cite{kkc87,KGSV,KFV,FKV0,FKV1,BronskiAn,BronskiNum,GarnierSIAM,GarnierWRM}, dispersive terms \cite{Burtsev}, nonlinear terms \cite{GarnierSIAM}, external force \cite{Gonz03}), (iii) random structure of corresponding coefficients (e.g. randomly distributed localized impurities \cite{kkc87,KGSV,KFV,FKV0,FKV1} or  finite-range inhomogeneities \cite{BronskiAn,BronskiNum,GarnierSIAM,Kalbermann,Gonz03}), and (iv) the statistics of the disorder potential like e.g. white or colored Gaussian potential \cite{GarnierWRM} or random step-wise process \cite{Knapp}. A noticeable share of the above mentioned papers is devoted to the soliton transmission through a 1D disordered segment in the framework of nonlinear Schr\"{o}dinger equation (NLSE). The latter is a ubiquitous nonlinear model appearing in many areas of contemporary physics. In the condensed matter physics NLSE occurs for example in the description of weakly nonlinear magnetization dynamics in ferromagnets with the easy-axis type anisotropy \cite{kik83}. It is also one of the main models in the nonlinear fiber optics and the nonlinear fiber arrays \cite{KA,dty05,fiberarrays}. A prominent features of the NLSE model is that in homogeneous (ordered) systems, this equation is completely integrable and possesses stable robust soliton solutions \cite{zs71,mnp84}.

Among the most common and widely explored sources of disorder are randomly distributed  point-like scatterers which emulate the random short-scale imperfection of a media \cite{kkc87,km89,BronskiNum}. When the concentration of scatterers is small, that is to say the mean distance between scatterers is much larger than the size of a soliton (i.e. the case of ``sparse scatterers''), the transmission of the soliton through a disordered segment can be considered as a
sequence of individual events of passing through a single scatterer.  After each act of scattering, the transmitted soliton acquires an abrupt shift of its position  as well as a change of both its energy and velocity (i.e. the scattering is generally inelastic). These changes influence the transmission characteristics of the soliton traversing through the total length of a sample, e.g. the time needed for the soliton to pass the entire segment (we call it a ``transmission time'' here). Assuming that the scattering is weak one can postulate that a soliton passing through a single scatterer, experiences only a slight change of its parameters, the measure of this ``slightness'' being the intensity of a scatterer. A single act of such a scattering was treated in detail in \cite{kkc87} within the framework of perturbation theory \cite{km89} (see also Ref. \cite{BronskiAn} where it was considered within another approach). When the perturbation method is applied, the resulting mapping mechanism for the soliton parameters is often called an ``equivalent particle approach'' due to the similarity of the resulting effect with the classical particle scattering \cite{BronskiNum}.

In this paper we study the statistics of the soliton transmission time $T$ through a 1D disordered segment in the framework of NLSE. More precisely, we address the delay for the transmission time of an incident soliton, $\Delta T$, occurring due to the presence of sparse random scatterers. Note, that often in the context of nonlinear optics the role of time is played by the spatial coordinate along the propagation of the optical beam, so for the so-called spatial solitons \cite{KA} this delay, $\Delta T$, should be endowed with a different physical meaning.

As mentioned above, the scattering itself is considered to be weak and a perturbative approach can be applied for the description of the soliton dynamics. In the first order with respect to weak scattering intensity, $\ve$, the time delay is related only to the shift in the soliton position after each scatterer. The corresponding total delay, $\Delta T$, is then proportional to the number of scatterers within the segment (and hence the segment length  $L$). On the other hand, the second order effects, i.e. those of the order of $\ve^2$, bring about the change of the soliton velocity, and their corresponding contribution is proportional to $L^2$. Therefore in the case of sufficiently long segment (the relevant inequalities will be given in the text), the latter contribution dominates, and the overall transmission delay should be calculated using the second order approximation with respect to the scatterer intensity.

We develop here a simple but yet robust mathematical formalism allowing one to obtain the mean value, $\left<\Delta T\right>$, and variance, $\sigma^2=\left<(\Delta T)^2\right>-(\left<\Delta T\right>)^2$, of the fluctuating transmission time delay. Some preliminary results concerning these first two moment have been published in \cite{KPGD}. But in the current paper not only we provide a detailed description of the method for the moments but also present the exact probability density, $\rho(\Delta T)$, and the probability $P(\Delta T<\Delta T_0)$ that the delay does not exceed a given value $\Delta T_0$. Such an approach has a certain merit on its own and can be used in the variety of similar random transmission problems where the velocity mapping between the two adjacent segments is given. It is pertinent to stress that in the current paper we consider the case of weak intensities of local inhomogeneities satisfying the inequality: $\ve^2 n \sim \ve^2 L \ll 1$, where $n$ is the typical number of scatterers. Thus the parameters of the transmitting soliton (i.e. its amplitude and velocity) change weakly, the measure of this weakness being the aforementioned parameter $\ve^2 n$. This case is somewhat opposite to that considered in Refs.\cite{BronskiAn,BronskiNum}, where the limit $\ve^2 L \to \infty$ was implied so that the limiting values of the soliton parameters differ significantly from their initial values.

The structure of the paper is as follows.  We start with brief reminder of the basic properties of a NLSE soliton,
consider soliton scattering on a single defect and recall the known expressions for the energy and the number  of emitted quasiparticles (subsection \ref{subsec:general}) used to obtain the formulas for velocity transformation  (subsection \ref{subsec:transf}). Our statistical model of randomly placed scatterers is introduced in section \ref{sec:model}, where we describe the exact (non-perturbative) method for the calculation of various ensemble averages. This method is then applied in Section \ref{sec:transm-many} to the problem of soliton transmission through a disordered segment: we analyze dynamics of the soliton passing trough the segment (subsection \ref{subsec:dynam}), and obtain general expressions for the mean transmission time delay and its variance (subsection \ref{subsec:MeanValue}). The limiting case of weak scatterers is the subject of subsection \ref{subsec:weakscat}: Here we obtain simple explicit formulas for the mean value and variance of the transmission time delay for both light and heavy solitons and estimate the length of the segment where second order contribution dominates over the first order one. The statistical properties of the delay time are studied in Section \ref{sec:delay}. In its first part (subsection \ref{subsec:gf}) we provide general formulas for various averages. The main result of this section is an explicit formula for the probability density function of transmission time delay obtained in subsection \ref{subsec:pdf}. In the last subsection  \ref{subsec:mandp} we present the probability that delay does not exceed a given value. The results obtained are summarized in Conclusion. Appendices \ref{sec:numenergy}-\ref{sec:equiv} contain the technical details of the calculations.

\section{Soliton transmission through a single weak scatterer}
\label{sec:transm-1}
\thispagestyle{headings}
\subsection{General remarks.}\label{subsec:general}
\thispagestyle{headings}
In this subsection we briefly recall some known results concerning the NLSE soliton transmission through the single weak point-like scatterer. Corresponding perturbed NLSE in the normalized dimensionless units reads \cite{kkc87}:
\begin{equation}
\label{NLSE-single}
 i u_t + u_{xx} + 2 |u|^2 u= u \ve \delta(x). \ \ \ \
 -\infty<x<\infty.
\end{equation}
Here  $u(x,t)$ is the complex field variable and the subscripts denote the partial derivatives with respect to
time $t$ and spatial coordinate $x$. The r.h.s. of this equation describes the influence of a single point scatterer with the intensity $\ve$ placed at origin. Let us mention again that in some systems (like e.g. for the spatial solitons in planar waveguides) the physical meaning of independent variables $t$ and $x$ may be different. Also the meaning of the field variable $u$ depends on the nature of the problem in hand: it can denote e.g. the deviation of the magnetization from the ``easy'' axis in ferromagnets or the envelope of the electromagnetic wave in optical fibers etc.

The unperturbed NLSE, i.e. when we set $\ve=0$ in Eq.(\ref{NLSE-single}), is completely integrable and possesses an infinite set of integrals of motion \cite{zs71}, including total energy $E$ and the total number of bound waves (quasiparticles) $N$. The simplest one-soliton solution of NLSE reads as \cite{mnp84}:

\begin{equation}
\label{one-sol}
    u_s(x,t)=ia\frac
    {\exp \left\{-i \left[
    \displaystyle{
    \frac{vx}{2}+\left(
    \frac{v^2}{4}-a^2\right)t}
    +\vp_0
    \right]\right\}}
    {\cosh\left[
    a(x+vt)-\vp_1
    \right]}
\end{equation}
and depends on four real parameters: $a$, $v$, $\vp_0$, $\vp_1$. We choose the soliton amplitude $a$ (which also characterizes its spatial size), and velocity $v$ to be positive. Such a selection corresponds to a soliton propagating in the negative direction of the $x$ axis.

Assuming the solution in the form of a single-soliton anstz, Eq. (\ref{one-sol}), the number of quasiparticles  $N_{s}$ bounded in the soliton (\ref{one-sol}) and  soliton energy  $E_{s}$ are given by

\begin{eqnarray}
        &N_s&= 2a, \\
  \label{Ints-0} E_s&=&\frac{av^2}{2}-\frac{2a^3}{3}.
\end{eqnarray}
\begin{figure}[t]
\includegraphics{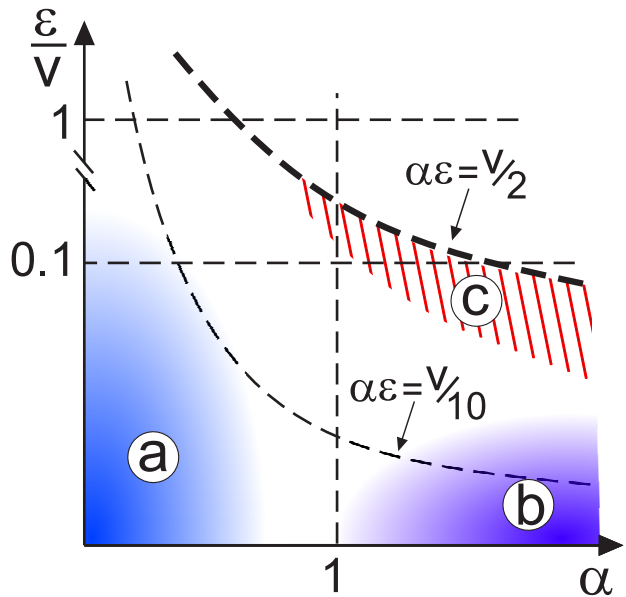}
  \caption{(Color online) The phase diagram indicating the different propagation regimes of soliton. The shadowed regions (a) and (b) correspond to fast light and fast heavy solitons correspondingly, while the dashed region (c) pertains to heavy slow solitons. The upper hyperbola, marks the boundary of the region of parameters where a soliton can pass over the impurity, see Eq.(\ref{Passing})}\label{Fig_1}
\end{figure}

In what follows we distinguish between the cases of ``light'' and ``heavy'' soliton. The particular type of the soliton is governed by dimensionless parameter:

\begin{equation}
\label{alpha}
    \alpha=\frac{2a}{v}.
\end{equation}
We name a soliton the amplitude of which is mush less than its velocity, $\alpha\ll 1$, a light soliton. Correspondingly heavy solitons are those for which the inverse inequality is satisfied, $\alpha\gg 1$.  This difference can be seen in Fig.\ref{Fig_1}: the region marked as (a) corresponds to the light solitons while regions (b) and (c) correspond to the heavy ones. The properties of light solitons are very close to those of a linear wave packet: almost all of its (positive) energy is contained in the first term in the r.h.s. of Eq.(\ref{Ints-0}) (i.e. in its kinetic energy). On the contrary, a heavy soliton behaves mostly like a classical particle and its (negative) energy is concentrated mainly in the second term in the r.h.s. of Eq.(\ref{Ints-0}). It means that the energy of the nonlinear interaction of the quasiparticles bound in the heavy soliton significantly exceeds their kinetic energy, and the opposite is true for the light solitons.



Consider now a soliton incident at $t\to-\infty$  from the right (i.e. from $x=+\infty$) and characterized by the amplitude $a$, velocity $v$ (or equivalently by the number of quasiparticles $N_s$ and energy $E_s$ (\ref{Ints-0})), and also by the phases $\vp_{0,1}$. As noted above, in this paper we will deal with the case of weak scatterers where the dimensionless scatterer intensity is small, and the ratio $\ve/v\ll 1$ is the main small parameter of the problem. Then one can resort to perturbation theory \cite{km89,km77} to describe the change of the soliton parameters after each act of scattering. The exhaustive perturbative study of the soliton transmission over a single delta-scatterer was carried out in Refs. [\onlinecite{kkc87}]. It was shown that in the case of attractive scatterer, $\ve<0$, the soliton becomes trapped in an effective potential well created by such an impurity and then experiences oscillatory motion in the vicinity of the scatterer. In the case of sufficiently strong repulsive scatterer, when the two inequalities $\ve>0$ and $v<2\ve \alpha$ are fulfilled, the soliton is always reflected by the scatterer (the region of parameters corresponds to the upper right sector in Fig. \ref{Fig_1}). The soliton can pass through the scatterer only in the case of comparatively weak repulsive interaction, and in our current study we assume that the following inequalities hold:

\begin{equation}\label{Passing}
0 < 2 \ve\alpha  < v.
\end{equation}

We distinguish two cases: i) a fast soliton, with parameters falling into the two marked regions, (a) and (b) in Fig.\ref{Fig_1}, where the strong version of inequality (\ref{Passing}) is valid:
\begin{equation}
\label{fast}
\quad \ve\alpha\ll v ,
\end{equation}
and ii) a slow soliton, where together with Eq.(\ref{Passing}) we have $\ve\alpha\sim  v$ (region (c) showed schematically by dashed lines in Fig.\ref{Fig_1}). As one can easily see, in the case of weak scatterers, $\ve \ll 1$, the light soliton is always fast, while the heavy soliton can be either fast (region (b) or slow (region (c)).

In the first order with respect to dimensionless scatterer intensity $\ve$, at $t\to+\infty$ the amplitude, velocity, the number of particles and energy of the soliton remain unchanged. The only changes are in the phases: $\vp_{0,1}'=\vp_{0,1}+\textit{O}$ $(\ve)$. Here and further on the primes will denote soliton parameters after scattering. These changes can be readily calculated \cite{kkc87} and as a result the position of the soliton center is asymptotically shifted back as compared to the unperturbed propagation of the soliton with the unchanged constant velocity $v$. The overall value of the backward coordinate shift (in the adiabatic approximation) is given by

\begin{equation}
\label{CenterShift-gen}
  d \approx 2\ve /v^2.
\end{equation}
For a slow soliton the additional shift of soliton position $d$ is of the order of its width $l_0 \sim 1/a$, and for a fast soliton this shift $d \ll l_0$.


Since in the first order in $\ve$ (adiabatic approximation) the amplitude $a$ and the velocity $v$ of the soliton do not change, its energy $E_s$ and the number of bound particles $N_s$ do not change either (see Eq.(\ref{Ints-0})) and there is no emission of quasiparticles from the soliton. But in the second order in $\ve$ (or rather $\ve/v$ as will be seen later), the solution at $t\to+\infty$ represents the transmitted soliton together with a number of quasilinear excitations carrying the number of waves $N$ and energy $E$. The amplitude and velocity of a scattered soliton do change now: \[ a'=a+\textit{O}\left(\ve^2/v^2\right),  \qquad v'=v+ O \left(\ve^2/v^2\right), \]
and so do the number of bound quasiparticles, $N_s'=N_s+ O \left(\ve^2/v^2\right)$, and the energy, $E_s'=E_s+O \left(\ve^2/v^2\right)$.
All the parameters describing the incident and scattered solution are related via the conservation laws for total number of quasiparticles and the total energy:

\begin{equation}
\label{conserv}
\begin{split}
  E_s &= E_s'+E, \\
  N_s &= N_s'+N.
\end{split}
\end{equation}

In the case of slow soliton the spectra of emitted quasiparticles can only be obtained numerically. However it appears that such a limit (when the position shift is of the order of the soliton width) is not a physical one. On the contrary, the case of fast soliton admits a detailed analytical description. Here the problem of emission of linear excitations can be solved perturbatively \cite{km89,BronskiAn}. In the second order of the perturbation theory the number of emitted quasiparticles and their energy have been calculated explicitly: the corresponding results in the cases of heavy and light soliton look rather different \cite{kkc87}. As was mentioned above, a heavy soliton behaves like a classical particle and the total number and energy of quasiparticles emitted by such a soliton are exponentially small:

\begin{equation}
\label{NewIntegrals-2}
    N\simeq\frac{2\pi\ve^2}{v}
    \left(\frac{a}{v}\right)^{9/2}
    {\rm e}^{-\displaystyle{\pi a/v}}, \qquad
    E\simeq\ve^2 v
    \left(\frac{a}{v}\right)^{11/2}
    {\rm e}^{-\displaystyle{\pi a/v}}, \ \ \ \ a\gg v.
\end{equation}
The behavior of a light soliton more or less mimics that of a linear wave packet.
In the leading approximation in $\alpha$, the expressions for the number and energy of quasiparticles emitted by a light soliton are as follows:

\begin{equation}
\label{NewIntegrals-3}
    N\simeq
    2a\left( \frac{\ve}{v}\right)^2, \qquad
    E\simeq\frac{\ve^2 a}{2}
    , \ \ \ \ a\ll v.
\end{equation}
\subsection{Transformations of the amplitude and velocity for the fast soliton. }\label{subsec:transf}
\thispagestyle{headings}
By virtue of Eqs. (\ref{conserv}) - (\ref{NewIntegrals-3}) one can express the energy change, $\delta E_s$, and the change of the number of quasiparticles, $\delta N_s$, via their emitted values. However the natural kinematic characteristics of the fast soliton are its amplitude and velocity. Now by means of the results of previous subsection we can obtain the mapping for the fast soliton amplitude and velocity during an act of scattering on a single defect. According to Eqs. (\ref{Ints-0}), (\ref{conserv}) the small changes of the soliton parameters after the scattering are given by (see Ref.\cite{BronskiAn}):

\begin{equation*}
\label{change-2}
    \delta N_s=-N=2 \, \delta a, \qquad \delta E_s = -E =
    \frac{v^2}{2} \, \delta a + a v \, \delta v- 2 a^2 \, \delta a.
\end{equation*}
These relations determine the relative change of soliton amplitude and velocity

\begin{equation}
\label{change-3}
\frac{\delta a}{a}=-\frac{N}{2a}, \ \ \ \
\frac{\delta v}{v}=-\frac{E}{av^2}+
\frac{N}{4a}-\frac{aN}{v^2},
\end{equation}
yielding the following evident velocity and amplitude transformation rules:

\begin{equation}
    v'=v\left[1-\frac{E}{av^2}+\frac{N}{4a}\left(1-\frac{4a^{2}}{v^{2}}\right)
    \right],
\end{equation}
\begin{equation}
    a'=a\left(1-\frac{N}{2a}\right),
\end{equation}
see Ref.\cite{BronskiNum}. A continuous version of these equations was obtained in Refs.\cite{BronskiAn,GarnierWRM}.

In the case of fast heavy soliton, where the inequalities are $\ve a\ll v^{2}$ and $a\gg v$, we can use the explicit expressions for the number $N$ and the energy $E$ of emitted quasiparticles (\ref{NewIntegrals-2}) mentioned in the previous Section \ref{subsec:general}. Here the last term in the second of Eqs. (\ref{change-3}) dominates and the transformation relations become:

\begin{equation}
\label{VelTransf-1}
    v' =v\left[1-2\pi\left(\frac{\ve}{v}\right)^2
    \left(\frac{a}{v}\right)^{11/2}{\rm e}^{-\displaystyle{\pi a/v}}\right],
     \qquad
    a' =a\left[1-\pi\left(\frac{\ve}{v}\right)^2
    \left(\frac{a}{v}\right)^{7/2}{\rm e}^{-\displaystyle{\pi a/v}}\right],
      \ \ \ a\gg v.
\end{equation}
The case of light soliton, $a \ll v$, is more subtle. In the leading approximation in $a/v$  from Eqs. (\ref{NewIntegrals-3}), due to the cancelation of contributions from energy and from number of emitted quasiparticles, the velocity does not change at all. Therefore the next terms in expansions of Eqs. (\ref{NewIntegrals-3}) should be taken into account. Corresponding calculations are performed in Appendix \ref{sec:numenergy}, and the eventual results are the following:

\begin{equation}
\label{NewIntegrals-4}
    N\simeq 2a\left(\frac{\ve}{v}\right)^{2}
    \left[1+\textit{O}\left(\frac{a^4}{v^{4}}\right)\right], \qquad
    E\simeq\frac{\ve^2 a}{2}
    \left[1-\frac{4a^2}{3v^{2}}+\textit{O}\left(\frac{a^4}{v^{4}}\right) \right], \ \ \ \ a\ll v.
\end{equation}

For velocity and amplitude transformation in the case of light soliton one obtains:

\begin{equation}
\label{VelTransf-2}
    v' =v\left(1-\frac{4}{3}\frac{\ve^2a^2}{v^4}
    \right),
     \ \ \ \ \ \
    a' =a\left(1-\frac{\ve^2}{v^2}\right),
     \ \ \ \ \ \ a\ll v.
\end{equation}

Generally, accounting for the results above the mapping rule can be represented in a unified way:

\begin{equation}
\label{Unified}
    a'=a\left[1-G(\ve,a,v)\right],\qquad
    v'=v\left[1-F(\ve,a,v)\right],
\end{equation}
where

\begin{equation}
\label{Unified-1}
      F(\ve,a,v)=\gamma \,\frac{a^2}{v^2}\, G(\ve, a,v),
\end{equation}
and

\begin{equation}
\label{Unified-2}
     \gamma=\left\{\begin{array}{ccc}
                                 \displaystyle{\frac{4}{3}}, & \ \ a\ll v, \\
 & \\ 2, & \ \ a\gg v ,                               \end{array}
     \right. \qquad
     G(\ve,a,v)=\left\{\begin{array}{lcc}
                                  \displaystyle{\left(\frac{\ve}{v}\right)^2}, \ \ & a\ll v ,\\ & \\
\pi\, \displaystyle{\left(\frac{\ve}{v}\right)^2
\left(\frac{a}{v}\right)^{7/2}}
{\rm e}^{-\displaystyle{\pi a/v}}, \ \ & a \gg v .                            \end{array}
     \right.
\end{equation}
One can see that in the case of light fast soliton, $a \ll v$, the relative change of the amplitude is determined only by a small parameter $(\ve/v)^2\ll 1$, while in the case of a heavy fast soliton, $a\gg v$, this change contains additional exponentially small factor. Also the velocity change in both cases contains an  additional parameter $(a/v)^2$. Thus the velocity change essentially exceeds the amplitude change for a heavy soliton, and is much smaller than the amplitude change for a light soliton. Let us note that the results very similar to Eqs. (\ref{Unified})-(\ref{Unified-2}) but for the different model of lengthy segment of random media were obtained in Ref.\cite{GarnierSIAM}.

\section{The model system of random scatterers}
\label{sec:model}
\thispagestyle{headings}
Let us now introduce disorder into the system and proceed from a single scatterer (defect) to a system of many scatterers with random positions and intensities. The NLSE for the system with many scatterers takes the form:

\begin{equation}
\label{NLSE-rand}
 i u_t + u_{xx} + 2 |u|^2 u= u \,\sum_k\ve_k \,\delta(x - x_k), \ \ \ \
k=1,2,3, ... , \ \ \ \ -\infty<x<\infty.
\end{equation}
The r.h.s. of Eq.(\ref{NLSE-rand}) describes the influence of the point scatterers with random
intensities $\ve_k$, placed at random positions $x_k$. We will consider the statistical properties of these quantities separately, starting with the intensities.

\subsection{The random intensities}\label{subsec:intens}
\thispagestyle{headings}
Let us consider the intensities of the defects as the mutually
independent random variables with the common probability density function,
$\tilde{\rho}(\ve)$,  and hence the same two first moments:
$\overline{\ve}$ and $\overline{\ve^2}\equiv \ve_0^2$ (the bar denotes averaging with the probability density $\tilde{\rho}(\ve)$). We assume here that the distribution of $\ve$ is not extremely exotic so that $\overline{\ve}$ and $\ve_0$ are of the same order of magnitude. The squared intensity of the $k$-th scatterer can be also characterized by a dimensionless parameter:

\begin{equation*}
\label{intensity}
\delta_k=\frac{\ve_k^2}{\ve_0^2}-1.
\end{equation*}
 Evidently, the first moment now $\overline{\delta_k}=0$, and all second moments are equal to the same value $\overline{\delta_k^2}\equiv \delta_0^2.$ The parameter $\delta_0^2$ is of the order of $(\Delta/\ve_0)^2$ where $\Delta$ is the width of the probability density function.
\subsection{Spatial distribution of scatterers}\label{subsec:spatial}
\thispagestyle{headings}
As for the spatial distribution of the defects, we will assume that they are  distributed \textit{independently and uniformly} within the segment $[0, L]$ with the mean distance $l$ between the adjacent scatterers. The number $n$ of
defects on the segment $[0,L]$ is random, and the probability $p_{n}$ to find exactly $n$ defects within the segment, is taken to be Poissonian:

\begin{equation}
    p_{n}=\frac{\Lambda^{n}}{n!}e^{-\Lambda}, \ \ \
    \sum_{n=0}^{\infty}p_{n}=1,
    \label{p-n}
\end{equation}
where $\Lambda=L/l$ is the average number of defects on the segment. Let us consider the probability density to find exactly $n$ scatterers at the points $x_k$, where we ordered the positions: $1\leq k\leq n$, $\ \ 0\leq x_n\leq x_{n-1}...\leq x_2\leq x_1\leq L$. It is convenient to introduce the new dimensionless variables:

\begin{eqnarray*}
\label{z}
    z_1=L^{-1}(L-x_1), \ \ \ \ z_k=L^{-1}(x_{k-1}-x_{k}), \ \ \ 2\leq k \leq n,
\end{eqnarray*}
see the scheme given in Fig.2. Now the aforementioned probability reads as:

\begin{eqnarray*}
    \rho_{n}(\{z\}_n)=n!
    \theta\left(1-\sum_{k=1}^{n}z_{k}\right)
    \prod_{k=1}^{n}\theta(z_{k}), \ \ \ \ \ \intop_0^\infty\rho_{n}(\{z\}_n)\prod_{k=1}^{n}dz_{k}=1, \ \ \ \ \{z\}_n=(z_1,z_2,...,z_n).
    \label{ro-1n}
\end{eqnarray*}

\begin{figure}[t!]
\includegraphics{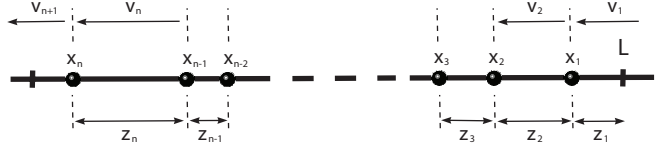}\label{Fig_2}
\caption{The scheme of a classical particle scattering on a disordered segment. }
\end{figure}

\subsection{The averages}\label{subsec:averages}
\thispagestyle{headings}
The dynamical quantities of the problem are described by various functions which depend on the number $n$,
set of positions $\{z\}$, and intensities $\{\delta\}$ of the scatterers. In this subsection we will present a simple and universal method of calculating the arbitrary averages over the intensities and Poisson-distributed positions of the scatterers. Again, the averaging with respect to intensities  of the scatterers is denoted by an overline:

\begin{equation*}\label{over}
    \overline{f(n,\{z\},\{\delta\})}=
    \int\rho_n(\{\delta\})
    f(n,\{z\},\{\delta\})
    D_n\{\delta\}, \ \ \ \
    \rho_n(\{\delta\}) =
    \rho(\delta_1)\ldots\rho(\delta_n),  \ \ \ \
    D_n\{\delta\}=d\delta_1\ldots d\delta_n.
\end{equation*}

Now, to distinguish between the intensity and position averaging the symbol $\langle \ldots \rangle_n$ will be used for canonical averaging with respect to positions of the scatterers:

\begin{equation*}\label{canon}
    \langle f(n,\{z\},\{\delta\}) \rangle_n=
    \int\rho_n(\{z\})
    f(n,\{z\},\{\delta\})
    D_n\{z\}, \ \ \ \
    D_n\{z\}=dz_1\ldots dz_n.
\end{equation*}
By ``canonical'' averaging we mean the averaging over all the realizations where exactly $n$ scatterers occur inside the segment $[0,L]$. Using the analogy with the notions of statistical mechanics we can introduce the ``grand canonical averaging'' over all the realizations having different number of scatterers with the weights given by Eq.(\ref{p-n}). The grand canonical averaging will be denoted by a single pair of angular brackets without the subscript $n$:

\begin{equation*}
\label{grandcan}
    \langle f(n,\{z\},\{\delta\})\rangle =\sum_n p_n \langle
    f(n,\{z\},\{\delta\}) \rangle_n.
\end{equation*}
Double angular brackets will denote the total averaging:

\begin{eqnarray}
\label{total}
    \langle \langle  f(n,\{z\},\{\delta\})\rangle \rangle =
    \overline{\langle f(n,\{z\},\{\delta\}) \rangle }=
\sum_n p_n \langle
    \overline{f(n,\{z\},\{\delta\})} \rangle_n.
\end{eqnarray}
In what follows it is useful to recall the expressions for the first two moments of the Poissonian distribution:
\begin{eqnarray*}
     \label{n-av}
     \langle n \rangle =\sum_{n=0}^\infty np_n=\Lambda; \ \ \ \
     \langle n^2 \rangle =\sum_{n=0}^\infty n^2p_n=\Lambda^2+\Lambda .
\end{eqnarray*}

We can also obtain the first two canonical moments of the distance
between the scatterers:

\begin{eqnarray}
     \label{z-av}
     \langle z_k \rangle_n =\int z_1\rho_{n}(\{z\})D_n\{z\}&=&
     \frac{1}{n+1}, \qquad
     \langle z_k^2 \rangle_n=\int z_1^2\rho_{n}(\{z\})D_n\{z\}=
     \frac{2}{(n+1)(n+2)},\nonumber\\
     \ \ \ \ \ \
     \langle z_j z_k \rangle_n&=&\int
     z_1z_2\rho_{n}(\{z\})D_n\{z\}=\frac{1}{(n+1)(n+2)},\ \ \
     \ \ \ \  j\neq k.
\end{eqnarray}
To obtain the latter quantities 
we have used the following trick. One starts with the well known integral
representation for the theta-function,
\begin{equation*}
\label{theta}
    \theta(u)=\frac{1}{2\pi i} \int_{C}\frac{d\kappa}{\kappa}
    e^{i\kappa u},
\end{equation*}
(the contour $C$ in the complex plane $\kappa$ is the line $\mathrm{Im} \,\kappa=-0$).
Then it is possible to interchange the order of the contour integration and the integration over all $z_k$.  Note that this trick comes handy for calculating arbitrary canonical averages, not only the first moments.
\section{The transmission through a disordered segment}
\label{sec:transm-many}
\subsection{Dynamics}\label{subsec:dynam}
In this section we consider the transmission time of a soliton passing through a disordered segment and calculate the mean value and variance of the transmission time shift.
It is important to mention that in the following analysis we account for the linear radiation only when applying the transformation formulas, Eqs. (\ref{Unified}). However we neglect all the secondary effects relevant to the multiply reflections of the radiation emitted previously by the soliton and the weak recurrent action of these waives on the overall soliton dynamics.

Let $v_1>0$ be the initial velocity of the soliton incident from the right on the segment $[0,L]$. In the absence of scatterers the transmission time would be $T_0=L/v_1$. The inclusion of random scatterers brings about two effects. First, the effective distance between the adjacent scatterers increases because of the backward shift of the soliton position. We denote such a backward position shift at the $k$-th scatterer as $d_k$. Second, a small deceleration of the soliton occurs after each act of scattering, which also increases the transmission time. Let $v_k>0$, $k \leq n$ be the velocity of soliton incident (from the right) on the $k$-th scatterer, and $v_{n+1}$ -- the velocity of the soliton after passing through the last $n$-th scatterer (note that the sequence of velocities $\{v_{k}\}$ monotonically decreases, $v_{k+1}<v_k$). Since the soliton propagates from right to left (i.e. from $+\infty$ to $-\infty$), we assume that $v_k$ denotes the absolute value of the velocity.

The total soliton transmission time $T_n$ through the disordered segment containing $n$ scatterers is:

\begin{equation}
\label{T}
    T_n=\frac{L-x_1}{v_1}+\frac{x_1-x_2+d_1}{v_2}+...
    \frac{x_{n-1}-x_n+d_{n-1}}{v_n}+\frac{x_n+d_n}{v_{n+1}}=T_0+\Delta T_{n1}+\Delta T_{n2},
\end{equation}
where the meaning of the quantities $\Delta T_{n1}$ and $\Delta T_{n2}$ is following.
By $\Delta T_{n1}$ we have designated the quantity:
\begin{equation}
\label{Times-one}
  \Delta T_{n1}= \sum_{k=1}^n \frac{d_k}{v_{k+1}}.
\end{equation}
So it is the contribution to the transmission time delay related to the backward position shift. This delay, $\Delta T_{n1}$, according to Eq. (\ref{CenterShift-gen}) can be expressed as:

\begin{equation}
\label{ShiftDelay}
    \Delta T_{n1}=\sum_{k=1}^{n}\frac{2\ve_k }{v_k^2\,v_{k+1}}.
\end{equation}

It follows from Eq.(\ref{VelTransf-1}) and Eq.(\ref{VelTransf-2}) that $(v_{k+1}-v_k)/v_k = (\ve_k/v_k)\,f(\alpha_k)$, so that in the leading approximation in $\ve$ one can  substitute $v_1$ for all the parameters $v_k$. Then we arrive at the expression:
\begin{equation}
\label{FirstOrder}
    \Delta T_{n1}\simeq \frac{2}{v_1^3}\sum_{k=1}^{n} \, \ve_k.
\end{equation}
and the final result for the first order contribution to the delay time is:
\begin{equation}
\label{FirstFinal}
    \overline{\Delta T_{n1}}=\frac{2n\overline{\ve} }{v_1^3}.
\end{equation}
The last term in the r.h.s.  of Eq.(\ref{T}), $\Delta T_{n2}$, is as follows:

\begin{equation}
\label{SecondOrder-0}
    \Delta T_{n2}= \sum_{k=1}^{n}x_k\left(
    \frac{1}{v_{k+1}}- \frac{1}{v_k}\right),
\end{equation}
so it describes the delay which occurs due to the soliton deceleration after each act of scattering. It is written in the most general form and as such describes the propagation time of an arbitrary classical particle  traversing through a disordered segment and obeying some given velocity transformation rules. Therefore the following analysis of this contribution in the current and following sections is applicable to any classical particle moving according to a given monotonically decreasing velocity mapping. Note that the velocity deceleration need not be small for the validity of the method -- only the positiveness of the velocity is important. We note that the intermediate calculations and  final results also look simpler in the general form rather than in the form of expansion in the limit of small $\ve/v$.  The latter limit is considered only in the subsection \ref{subsec:weakscat}, where we obtain and analyze the mean value and the variance of the transmission time delay for a fast soliton.

In the dimensionless variables Eq.(\ref{SecondOrder-0}) can be recast as:

\begin{equation}
\label{SecondOrder}
    \Delta T_{n2}= L\,
    \sum_{k=1}^{n}\left(
    \frac{1}{v_{k+1}}- \frac{1}{v_k}\right)
    \left(1-\sum_{j=1}^{k}z_j
    \right).
\end{equation}
The statistical analysis of the shift $\Delta T_{n2}$ is more complicated than that of $\Delta T_{n1}$  and the rest of the paper is dedicated to the former. As we will see below this contribution dominates for long enough segments with the large average number of scatterers. Therefore in what follows we will omit the subscript ``2" and write simply $\Delta T_n$ implying the quantity $\Delta T_{n2}$ unless specified otherwise.

The corresponding delay depends on the number $n$ of the scatterers and on their realization, i.e. on the two sets of parameters $\{z\}\equiv\{z_1,\ldots ,z_n\}$ and $\{\delta\}\equiv\{\delta_1 , \ldots \delta_n\}$ (the latter enters through the velocities $v_k$). Indeed according to Eqs. (\ref{Unified}), the change of velocities is described by the set of recurrent relations:
\begin{equation}
\label{recurrency}
    v_{k}=v_{k-1}\big[1-F(\ve_{k-1},a_{k-1},v_{k-1})\big), \ \ \ \ \ \
    a_{k}=a_{k-1}\big[1-G(\ve_{k-1},a_{k-1},v_{k-1})\big],
\end{equation}
which actually gives $v_{k}$ as a function of an input velocity $v_1$, input amplitude $a_1$ and fluctuations $\delta_l$ of intensities $\ve_l$ of all scatterers $l=1,2,...,k-1$ preceding (from the right) the scatterer with the number $k$,

\begin{equation}
\label{VelTransf}
    v_k=\Phi
    (v_1,a_1;\delta_1,\delta_2,...,\delta_{k-1}).
\end{equation}

We emphasize again that in the case considered, the soliton behaves as a classical particle that moves with a constant velocity between the scatterers. The corresponding dimensionless transmission time shift,

\begin{eqnarray}
\label{tau-1}
    \tau_{n}(\{z\},\{\delta\})\equiv\frac{\Delta T_n}{T_0}=
    \mu_{n,1}-\sum_{k=1}^{n}\mu_{n,k}z_{k},
\end{eqnarray}
is expressed via the natural dynamic variables:

\begin{equation}
\label{mu}
    \mu_{n,k}=\frac{v_{1}}
    {v_{n+1}}-\frac{v_{1}}{v_{k}}>0, \ \ \ \ k=1,2,\ldots,n,
\end{equation}
which are nothing else but dimensionless shifts of the inverse velocities after passing through the last $n-k+1$ scatterers. Thus we see that the fluctuations of intensities and positions of the scatterers in Eq. (\ref{tau-1}) are decoupled: accounting for Eq. (\ref{VelTransf}) the former enter only through the variables $\mu_{n,k}$ while the geometric disorder enters directly through the dimensionless distances $z_k$ between the adjacent scatterers.


Because the velocities form a monotonically decreasing sequence, the sequence $\mu_{n,k}$ also decreases: $\mu_{n,k+1}<\mu_{n,k}$. From Eq. (\ref{tau-1}) it follows that the shift $\tau_{n}$ does not exceed the value $\mu_{n,1}$:

\begin{equation*}
\label{tau-2}
    0\leq\tau_{n}(\{z\},\{\delta\})\leq\mu_{n,1}.
\end{equation*}
The maximum of the transmission time occurs in the configuration where all the scatterers are concentrated
at the point of incidence (all $x_k=L$)
so that the soliton always moves with the minimal
velocity $v_{n+1}$. The minimal shift equals zero and corresponds to
the opposite configuration where all the scatterers are concentrated at
the farther end of the segment (all $x_k=0$) and the soliton always moves with its initial
(maximal) velocity $v_{1}$.
\subsection{The mean value and the variance of the transmission time shift}
\label{subsec:MeanValue}
The statistical properties of the dimensionless transmission time shift, Eq.(\ref{tau-1}), are the main subject of our paper. We will show that within the framework of the classical ``particle'' model one can construct a complete statistical description of the transmission time delay and obtain a general expression for its probability density function. This problem will be considered in the next Section \ref{sec:delay}.

However, sometimes for practical applications one may want to  know only the first two moments of the delay. In this subsection we will calculate separately the mean shift of the transmission time and its variance. To calculate the variance, besides Eq.(\ref{tau-1}) we will also need the expression for the square of the shift in terms of $\mu_{n,k}$ and $z_k$:

\begin{eqnarray}
\label{tau-3}
    \tau_{n}^2(\{z\},\{\delta\})=\mu_{n,1}^2-2\mu_{n,1}\sum_{k=1}^n
    \mu_{n,k}z_k+
    \sum_{k=1}^n \mu_{n,k}^2 z_k^2+\sum_{k\neq m}^n
    \mu_{n,k} \mu_{n,m}z_k z_m.
\end{eqnarray}

The canonical averaging of
Eqs. (\ref{tau-1}),
(\ref{tau-3}),
$$
\tau_n^{(1)}=\langle \overline{\tau_n(\{z\},\{\delta\})} \rangle_n, \qquad
\tau_n^{(2)}=\langle \overline{\tau_n^2(\{z\},\{\delta\})} \rangle_n,
$$
by virtue of Eqs. (\ref{z-av}) leads  to the following general
expressions:

\begin{eqnarray}
\label{means}
    \tau_n^{(1)}&=& \overline{\mu_{n,1}}-\frac{1}{n+1}\sum_{k=1}^n
    \overline{\mu_{n,k}} \, ,\\
    \tau_n^{(2)}&=&
    \overline{\mu_{n,1}^2}-\frac{2}{n+1}
    \displaystyle{\sum_{k=1}^n
    \overline{\mu_{n,1}\mu_{n,k}}}+
    \frac{1}
    {(n+1)(n+2)}\left[
    \displaystyle{\sum_{k=1}^n\overline{ \mu_{n,k}^2} +
    \overline{ \left( \sum_{k=1}^n \mu_{n,k} \right)^2 }}
    \right].
    \label{means2}
\end{eqnarray}

Eventually the total averages can be obtained by applying the averaging formula, Eq. (\ref{total}):

\begin{eqnarray}
\label{TauTotal}
    \tau^{(1)}&=&\langle \langle \tau_{n}(\{z\},\{\delta\}) \rangle \rangle
=\sum_{n=0}^{\infty}p_n\tau_n^{(1)}, \nonumber\\
\tau^{(2)}&=& \langle \langle \tau_{n}^2(\{z\},\{\delta\}) \rangle \rangle
=\sum_{n=0}^{\infty}p_n\tau_n^{(2)}.
\end{eqnarray}

\subsection{Weak scatterers limit}\label{subsec:weakscat}
Eqs. (\ref{means}),(\ref{means2}) were obtained for the general type of velocity mapping:

\begin{equation}
\label{mapping}
    v_{j+1}=v_j\left(1-F(v_j)\right).
\end{equation}
Here we consider the weak scattering limit where the parameter

\begin{equation}
\label{smallparameter-1}
    \zeta=\frac{v_1-v_2}{v_1}\equiv F(v_1)
\end{equation}
is small. More precisely, to avoid the error accumulation it should be much less than inverse number of scatterers on the segment

\begin{equation}
\label{weakscatteringlimit}
    n\zeta\ll 1.
\end{equation}
This equation serves as definition of the weak scattering limit. Turning to the specific problem of the soliton scattering, the function $F(v_1)$ coincides with the function $F(\ve_0,a_1,v_1)$ from Eq. (\ref{Unified-1}).

In the first order in $\zeta$ the mapping relations defined by Eqs. (\ref{recurrency}), (\ref{mu}) become much simpler.  For the fast soliton they read:

\begin{eqnarray}
    \label{weak}
    &&v_k=v_1\left[1-\zeta
    \left(k-1+\sum_{m=1}^{k-1}\delta_m\right)\right],\nonumber\\
    &&\mu_{n,k}=\zeta
    \left(n-k+1+\sum_{m=k}^{n}\delta_m\right).
\end{eqnarray}
Note that the velocity mapping decouples from the amplitude mapping in this approximation. The canonical averages can  now be straightforwardly calculated:
\begin{eqnarray}
\label{tau-means}
    \tau_n^{(1)}=\zeta\, \frac{n}{2}, \qquad
    \tau_n^{(2)}=\zeta^2\left(\frac{n}{2}\right)^2\left[
    1+\frac{1}{3n}(1+4\delta_0^2)
    \right].
\end{eqnarray}

 One notes that in the case of large number of scatterers $n$, $ \zeta\ll n^{-1} \ll 1$, both dimensionless shift of the transmission time,
$\tau_n^{(1)}$, and its relative standard deviation,
\begin{equation}\label{var1}
\sqrt{\frac{\tau_n^{(2)}}{\left( \tau_n^{(1)}\right)^2}-1
}=\sqrt{\frac{1+4\delta_0^2}{3n}},
\end{equation}
are small. The smallness of fluctuations is provided by large number of scatterers $n$ only, while the small deviations from the unperturbed transmission time require both the weak strengths of the scatterers $\varepsilon$ and/or the large initial velocity $v_1$ (recall that $\ve_0/v_1 \ll 1/\sqrt{n} \ll 1$).

After the next averaging over the number of scatterers, Eq.(\ref{TauTotal}), we obtain:

\begin{eqnarray}\label{results}
\tau^{(1)}=\sum_{n=0}^\infty p_n \tau_n^{(1)}=\frac{\zeta\Lambda }{2}, \qquad  \tau^{(2)}=\sum_{n=0}^\infty p_n
\tau_n^{(2)}=\left(\frac{\zeta\Lambda }{2}\right)^2+\frac{\zeta^2\Lambda}{3}
(1+\delta_0^2).
\end{eqnarray}
These results are qualitatively the same as those for the canonical ensemble, cf. Eqs.(\ref{tau-means}). The difference is that here the average number of scatterers, $\Lambda=L/l$, stands for $n$ from Eqs.(\ref{tau-means}), and some of the numerical
coefficients have changed. Finally, going back to the dimensional variables we obtain the weak scattering expressions for the mean transmission time shift $\left<\Delta T\right>,$ and its standard deviation $\delta T=\big( \left<\left(\Delta T\right)^2\right>-\left<\Delta T\right>^2 \big)^{1/2}$:

\begin{equation}\label{T-can}
\langle \Delta T \rangle_n=\frac{n\zeta}{2}T_0, \qquad \delta T_n=\langle \Delta
T \rangle_n {\sqrt\frac{1+4\delta_0^2}{3n}},
\end{equation}
for the canonical ensemble, and

\begin{equation}\label{T-gcan}
\langle \langle \Delta T \rangle \rangle=\frac{L\zeta}{2\, l}T_0,
 \qquad \delta
T=\langle \langle \Delta T \rangle \rangle \sqrt{\frac{4l(1+\delta_0^2)}
{3L}},
\end{equation}
for the grand canonical ensemble. The expressions above present an improved version of those given in our earlier Ref. \cite{KPGD} where the definition of parameter $\zeta$ for the case of a light soliton was erroneous.

 Note that the values of small parameter $\zeta$ for light and heavy solitons are different. Therefore  within the canonical ensemble, the mean shift of the transmission time is:

\begin{equation}
\label{shift}
    \langle \Delta T \rangle_n=\left\{
    \begin{array}{cccc}
      \displaystyle{\frac{2}{3}\frac{nL}{v}\frac{\ve_0^2}{v^2}
      \left(\frac{a}{v}\right)^2}, &  & & \textrm{light soliton} \ \ a\ll v, \\
        &  & &   \\
      \pi\displaystyle{
                 \frac{nL}{v} \frac{\ve_0^2}{v^2}
                 \left(\frac{a}{v}\right)^{11/2}
                 \exp\left(-\pi\frac{a}{v}\right)},&  & & \textrm{heavy soliton} \ \ a\gg v.
    \end{array}
    \right.
\end{equation}
Here and in the two following Eqs. (\ref{crossover}) and (\ref{FinalShift}), we set for simplicity $a\equiv a_{1}$ and $v\equiv v_{1}$ as the designations for the input amplitude and velocity values.
The results for grand canonical ensemble are obtained by replacing the number of scatterers $n$ by its mean value $L/l$. The standard deviation is then found from Eqs. (\ref{T-can}), (\ref{T-gcan}).


Recall now that the results above are relevant for the second order contribution in $\ve$, $\Delta T_{n2}$ (which is due to the velocity shifts and deceleration of the soliton) to the transmission time delay which coexists with the first order contribution $T_{n1}$, Eq. (\ref{FirstFinal}), (which arises due to the position shifts). Each of the two is dominant in its own interval of the segment lengthes. The crossover length, $L_c$, is defined by the expression:

\begin{equation}
\label{crossover}
    L_c \sim \left\{
    \begin{array}{cccc}
      \displaystyle{
      \frac{1}{\ve_0}
      \left(\frac{v}{a}\right)^2
      }, &  & &
      \textrm{light soliton} \ \ a\ll v, \\
        &  & &   \\
      \displaystyle{
                 \frac{1}{\ve_0}
                 \left(\frac{v}{a}\right)^{11/2}
                 \exp\left(\pi\frac{a}{v}\right)
      },&  & &
      \textrm{heavy soliton} \ \ a\gg v ,
    \end{array}
    \right.
\end{equation}
 For short segments, $L\ll L_c$, the transmission time delay is mostly determined by the first order contribution due to backward shift of the soliton center after each scattering, see Eq.(\ref{FirstFinal}). On the contrary, for the long segments, with $ L\gg L_c$, the velocity deceleration plays crucial role and the delay is described by Eq. (\ref{shift}). Finally, combining the results in the two aforementioned regimes we can write the mean transmission time delay for canonical ensemble as a general expression:

\begin{equation}
\label{FinalShift}
    \langle \Delta T \rangle_n=\left\{
    \begin{array}{cccc}
      \mathrm{max} \left[\displaystyle{\frac{2n\overline{\ve}}{v^3}, \ \ \
      \frac{2\,nL\ve_0^2}{3\,v^3}\left(\frac{a}{v}\right)^2}\right], &  & & \textrm{light soliton} \ \ a\ll v, \\
        &  & &  \\
      \mathrm{max} \left[\displaystyle{\frac{2n\overline{\ve}}{v^3}, \ \ \
      \frac{\pi\, nL\ve_0^2}{v^3}\left(\frac{a}{v}\right)^{11/2}}
      \displaystyle{\exp\left(-\pi\frac{a}{v}\right)}\right],&  & & \textrm{heavy soliton} \ \ a\gg v.
    \end{array}
    \right.
\end{equation}

\section{Statistical properties of the transmission time delay}
\label{sec:delay}
\subsection{General formulas}\label{subsec:gf}
In the previous section the two first moments of the time shift were
calculated both in general situation and in the weak scatterer approximation.
The exact results were applicable to any classical ``particle'' evolving according the prescribed velocity mapping between the scatterers.  Here we perform the analysis a step further and obtain the probability density function for the soliton transmission shift of a scattered ``particle''.

The latter can be written as a grand canonical averaging of the corresponding delta function:
\begin{equation}
\label{P-tot-1}
    \rho(\tau)=
    \langle \langle \delta\big(\tau-\tau_n[\{z\},\{\delta\}]\big) \rangle \rangle.
\end{equation}
Expressed in terms of canonical averages this probability density is:

\begin{equation}
    \rho(\tau)=
\sum_{n=0}^{\infty}p_n \rho_n(\tau).
    \label{P-tot-2}
\end{equation}
Here $\rho_0(\tau)= \delta(\tau),$ and for all $n\geq 1$

\begin{equation}
\label{P-tot-3}
    \rho_n(\tau) = \overline{\rho(\tau;n,\{\delta\})},
\end{equation}
with the partial probability density

\begin{equation*}
\label{P-tot-4}
    \rho(\tau;n,\{\delta\})=
    \langle \delta(\tau-\tau_n(\{z\},\{\delta\})) \rangle_n.
\end{equation*}
The probability density will be explicitly calculated in the following subsections.
\subsection{Probability density function}\label{subsec:pdf}
The intensities of the scatterers enter the partial probability density,
$\rho(\tau;n,\{\delta\})$, through the set of parameters $\mu_{n,k}$. This dependence should be taken into account explicitly only during the averaging over all realizations of the set $\{\delta\}$.
In what follows we will consider the canonical configurational averaging (i.e averaging over the positions of the scatterers, $z_k$) and omit the symbol $\{\delta\}$ in the argument of the partial probability density.

For $n=1$ after straightforward integration over $z_1$ one gets:

\begin{equation}
\label{Pi-one}
    \rho(\tau;1)=\frac{1}{\mu_{1,1}}
    \theta(\mu_{1,1}-\tau).
\end{equation}
For all $n\geq 2$ we also start with the integration over $z_{1}$.  Due to the presence of delta function, the only
point contributing to the integral is

\begin{equation*}
\label{z(z)}
    z_{1}= z_{1}(\{z\})\equiv 1-\frac{\tau}{\mu_{n,1}}-
    \frac{1}{\mu_{n,1}}\sum_{k=2}^{n}z_{k}\mu_{n,k},
\end{equation*}
which leads to the result:

\begin{eqnarray}
\label{P-part-2}
    \rho(\tau;n)  =  \frac{n!}{\mu_{n1}}
    \int_{0}^{\infty}...\int_{0}^{\infty}
    \theta(z_{1}\{z\}) \ \
   \theta\left(1-z_{1}\{z\}-\sum_{k=2}^{n}z_{k}\right)\prod_{k=2}^{n}dz_{k}.
\end{eqnarray}
Using the integral representation
for the $\theta-$function from subsection \ref{subsec:averages}, we  can rewrite the partial probability density as:

\begin{eqnarray*}
\label{P-part-3}
    \rho(\tau;n) =&& \frac{n!}{\mu_{n,1}}
    \frac{1}{(2\pi i)^{2}}
    \int_{C_{1}}\frac{d\kappa_{1}}{\kappa_{1}}
    \int_{C_{2}}\frac{d\kappa_{2}}{\kappa_{2}}
    \exp\left[i\kappa_{1}\left( 1-\frac{\tau}{\mu_{n,1}}\right)
    +i\kappa_{2}\frac{\tau}{\mu_{n,1}}\right]\times\nonumber\\
    &&\prod_{k=2}^{n}
    \left\{  \int_{0}^{\infty}dz_{k}
    \exp
    \left[
    -z_{k}\left(i\kappa_{1}\frac{\mu_{n,k}}{\mu_{n,1}}
    +i\kappa_{2}
    \left( 1-\frac{\mu_{n,k}}{\mu_{n,1}}
    \right)
    \right)
    \right]
    \right\}.
 \end{eqnarray*}
All integrals over $\{z\}$ converge because the sequence $\mu_{ni}$ decreases.  After integration we obtain:

\begin{eqnarray*}
\label{P-part-4}
    \rho(\tau;n) = \frac{n!}{\mu_{n,1}}
    \frac{1}{(2\pi i)^{2}}
    \int_{C_{1}}\frac{d\kappa_{1}}{\kappa_{1}}
    \int_{C_{2}}\frac{d\kappa_{2}}{\kappa_{2}}
    \exp\left[i\kappa_{1}\left( 1-\frac{\tau}{\mu_{n,1}}\right)
    +i\kappa_{2}\frac{\tau}{\mu_{n,1}}\right]
    \prod_{k=2}^{n}
    \left[
    i\kappa_{1}\frac{\mu_{n,k}}{\mu_{n,1}}
    +i\kappa_{2}
    \left( 1-\frac{\mu_{n,k}}{\mu_{n,1}}
    \right)
    \right]^{-1}.
 \end{eqnarray*}

The next step is the integration over $\kappa_{1}$ and $\kappa_{2}$. We start with integration over $\kappa_{1}$. The integrand has simple poles at points:
$$
\kappa_{1}=\kappa_{2}\left(1-\frac{\mu_{n,1}}{\mu_{n,k}}\right),
k=1,2,\ldots, n ,
$$
which lie at the upper half plane of $\kappa_{1}$. If $\tau>\mu_{n,1}$, we close contour $C_{1}$ in the lower
half-plane and the integral is equal to zero. In the opposite case, $\tau<\mu_{n,1}$, the integral is
proportional to the sum of residues in all the poles:

\begin{eqnarray*}
\label{P-part-5}
    \rho(\tau;n)  =
    \frac{
    \theta(\mu_{n,1}-\tau)
    n!
    }
    {2\pi{\displaystyle \prod_{k=1}^{n}(i\mu_{n,k})}
    }
    \sum_{j=1}^{n}
    \left[
    {\displaystyle
    \prod_{k=1,k\neq j}^{n}
    \frac{\mu_{n,j}-\mu_{n,k}}{\mu_{n,j}\ \ \mu_{n,k}}
    }
    \right]^{-1}
    \int_{C_{2}}\frac{d\kappa_{2}}{(\kappa_{2})^{n}}
    \exp\left\{
    {\displaystyle
    \frac{i\kappa_{2}}{\mu_{n,j}}
    \left[
    \tau-
    \left(
    \mu_{n1}-\mu_{n,j}
    \right)
    \right]}
    \right\},
    \ \ n \geq 2.
 \end{eqnarray*}
The integral above differs from zero only for $\tau>\left(\mu_{n,1}-\mu_{n,j}\right).$ In this case one should
calculate the contribution from a single pole $\kappa_{2}=0$ of the order $n$. The result is:

\begin{eqnarray}
\label{P-part-7}
    \rho(\tau;n)  =
    \sum_{j=1}^{n}
    \frac{f_{j}(\tau;n)}
    {{\displaystyle
    \prod_{k=1, k\neq j}^{n}(\mu_{n,j}-\mu_{n,k})}
    }, \ \ \ n\geq 2,
 \end{eqnarray}
where
\begin{eqnarray}
\label{f}
    f_{j}(\tau;n)=\theta(\mu_{n,1}-\tau)
    \theta(\tau-(\mu_{n,1}-\mu_{n,j}))
    \frac
    {n[\tau-(\mu_{n,1}-\mu_{n,j})]^{n-1}}
    {\mu_{n,j}}, \qquad \int_{0}^{\mu_{n,1}}f_{j}(\tau;n)d\tau=\mu_{n,j}^{n-1}.
 \end{eqnarray}

The specific structure of Eq. (\ref{P-part-7}) enables us to represent it as a ratio of two determinants:

\begin{eqnarray}
\label{P-part-8}
    \rho(\tau;n) =\frac{W_{n}^{f}\{\mu_{n}\}}{W_{n}\{\mu_{n}\}},
 \end{eqnarray}
where $W_{n}\{\mu_{n}\}$ is  the Vandermonde  determinant based on the powers of $\mu_{n,k}$:

\begin{equation}
\label{vdM-1}
    W_{n}\{\mu_{n}\}=
    \left|\begin{array}{cccc}
    \mu_{n,1}^{n-1} & \mu_{n,2}^{n-1} & \ldots &
    \mu_{n,n}^{n-1} \\
    \mu_{n,1}^{n-2} & \mu_{n,2}^{n-2} & \ldots & \mu_{n,n}^{n-2}\\
    \ldots & \ldots & \ldots & \ldots \\
    \mu_{n,1} & \mu_{n,2} & \ldots  & \mu_{n,n}\\
    1 & 1 & \ldots & 1 \\
    \end{array}\right|,
\end{equation}
and  $W_{n}^{f}\{\mu_{n}\}$ is obtained from the expression above by replacing the first line with the corresponding functions $f_k$ given by Eq.(\ref{f}):

\begin{equation}
\label{vdM-2}
    W_{n}^{f}\{\mu_{n}\}=
    \left|\begin{array}{cccc}
    f_1 & f_2 & \ldots  &
    f_n \\
    \mu_{n,1}^{n-2} & \mu_{n,2}^{n-2} & \ldots & \mu_{n,n}^{n-2}\\
    \ldots & \ldots & \ldots & \ldots \\
    \mu_{n,1} & \mu_{n,2} & \ldots & \mu_{n,n}\\
    1 & 1 & \ldots & 1 \\
    \end{array}\right|.
\end{equation}
Indeed, ordering the differences in the denominators in the l.h.s.
of Eq.(\ref{P-part-8}), so that all of them are positive, and
summing up all terms, we get the common denominator
$$
\prod_{1\leq k<j\leq n}\left( \mu_{n,k}-\mu_{n,j}\right)=
W_{n}\{\mu_{n}\},
$$
while the corresponding numerator is nothing but the expansion of
$W_{n}^{f}\{\mu_{n}\}$ with respect to its first line. Thus
the delay probability density is defined by Eqs. (\ref{P-tot-2}),
 (\ref{P-tot-3}) with partial probability density function $\rho(\tau;n,\{\delta\})$ given by Eq.(\ref{P-part-7}).

We recall that the scaled transmission time delay, $\tau$, enters into this formula through the function $f$, see Eq. (\ref{f}), while the realization of scatterer intensities, $\{\delta\}$, enters through the variables $\mu$ which are determined via soliton velocities $v_k$ obeying the velocity transformation,  Eq. (\ref{VelTransf}). Eqs. (\ref{p-n}) and (\ref{f}) automatically provide normalization condition of the probability density of the transmission time shift:

\begin{equation*}
\label{Norm-2}
    \int_{0}^{\infty}\rho(\tau)d\tau=1.
\end{equation*}

In particular case of identical weak scatterers, $0<\ve_j=\ve\ll 1$, the results obtained look essentially simpler. Indeed, in this case according to Eq. (\ref{weak}) we get:

\begin{equation}
    \label{weak-1}
    \mu_{n,k}=\zeta(n-k+1),
\end{equation}
and in terms of the scaled time shift, $\tilde{\tau}=\tau/\zeta$, the partial probability density $\rho(\tilde{\tau},n)=\zeta\rho(\tau,n)$ reads:

\begin{equation}
\label{partial-7}
     \rho(\tau;n)=\left\{\begin{array}{lcc}
                                  \delta(\tilde{\tau}), \ \ & n=0 ,\\ & \\
\theta(1-\tilde{\tau}), \ \ & n=1,\\&\\
 \theta(n-\tilde{\tau})\displaystyle{\sum_{j=1}^{n}}
 \theta(\tilde{\tau}-j+1)P(\tilde{\tau};n,j), \ \ & n\geq 2.                       \end{array}
     \right.
\end{equation}
Here all $P(\tilde{\tau};n,j)$ are $(n-1)-$power polynomial functions of $\tilde{\tau}$:

\begin{equation}
\label{polynoms}
     P(\tilde{\tau};n,j)=
    \frac{n}{n-j+1}
    \frac{(\tilde{\tau}-j+1)^{n-1}}
    {{\displaystyle
    \prod_{k=1, k\neq j}^{n}(k-j)}
    }.
\end{equation}

All partial PDFs, Eqs. (\ref{partial-7}), are universal. The fact that we consider either light or heavy soliton, as well as the dependence on a particular set of initial parameters $a_1$, $v_1$ and $\ve_0$, is reflected only in a particular value of scaling parameter $\zeta$, Eq. (\ref{smallparameter-1}). Therefore in the case of identical weak scatterers the canonical PDFs given by Eqs. (\ref{partial-7}), not only describe the statistical properties of the time shift of an NLSE soliton but are also applicable to any classical object with the velocity transformation law

\begin{equation}
\label{transformation}
    v_{j+1}=v_j(1-\zeta).
\end{equation}

\begin{figure}[b!]
\includegraphics{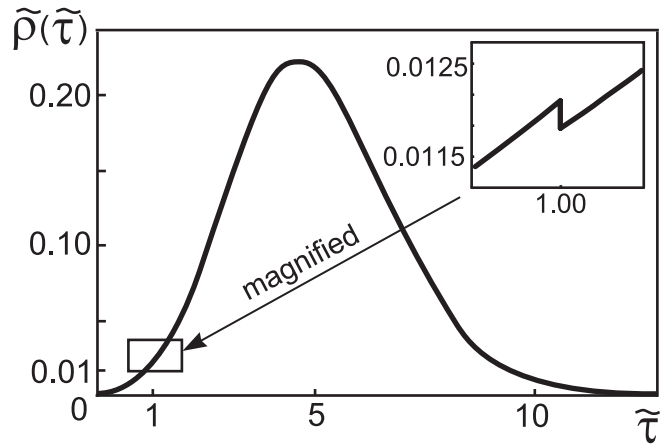}
  \caption{Truncated grand canonical PDF for the scaled transmission time delay. The jump of the PDF at $\tilde{\tau} = 1$ (magnified) is shown in the inset.}
  \label{Fig_3}
\end{figure}

With the growth of the partition number $n$, the singularities in PDF become weaker. The zeroth term, $n=0$, contains a delta-singularity at the origin, the first term has a jump at $\tilde{\tau}=1$. All higher partial PDFs, $\rho(\tilde{\tau};n)$, with numbers $n=2,3,...$ are continuous together with their first $n-2$ derivatives. However already $n-1-$th derivatives have a jump at the points $\tilde{\tau}=1,2,...,n$. For the points $\tilde{\tau}=1,2,...,n-1$ this statement is evident (see Eqs. (\ref{partial-7}), (\ref{polynoms})). For the point $\tilde{\tau}=n$ it immediately follows from the representation given by Eqs. (\ref{P-part-8}), (\ref{vdM-1}).

In the general case, performing the summation of canonical PDFs, Eq.(\ref{P-part-8}), with the Poissonian weights given by Eq.(\ref{p-n}), we arrive at the grand canonical PDF,  $\rho(\tau)$, for the transmission time. In the same approximation of identical weak scatterers the grand canonical PDF, $\rho(\tilde{\tau})$, (in scaled variable $\tilde{\tau}$) depends on a sole dimensionless parameter $\Lambda=L/l$ which is nothing but the mean number of scatterers in the segment. In  Fig.\ref{Fig_3} we displayed the truncated grand canonical PDF,

\begin{equation}
\label{truncatedPDF}
    \tilde{\rho}(\tilde{\tau})=\displaystyle{\sum_{n=1}^{\infty}}p_n\rho(\tilde{\tau},n)=
    \rho(\tilde{\tau})-e^{-\Lambda}\delta(\tilde{\tau}),
\end{equation}
for $\Lambda=10$, i.e. we extracted the singular contribution at the origin. The truncated (regular) PDF is continuous on a whole semiaxis $\tilde{\tau}>0$ except for the point $\tilde{\tau}=1$. At this point the PDF has a jump with the magnitude which according to Eq. (\ref{p-n}) is equal to $0.000454$. This small jump is shown in the inset panel of Fig. \ref{Fig_3}.

The results above allows one to gather all statistical information about the time delay in the disordered segment. One can verify directly (see Appendix \ref{sec:Mom}) that the expression for PDF given by Eq.(\ref{P-part-8}) leads to the same expressions for the two first moments of time delay as was given by Eq.(\ref{means}) before.

\subsection{The cumulative distribution function}\label{subsec:mandp}

Another important quantity of interest is the cumulative distribution function, i.e. the probability $P(\tau_0)$ that the dimensionless transmission time shift, $\tau=\Delta T/T_0=v_1\Delta T/L$, does not exceed a fixed value $\tau_0$.
This probability,

\begin{equation*}
\label{Prob}
    P(\tau_0)\equiv P(\tau\leq\tau_0)=\langle \langle \theta(\tau_0-\tau) \rangle \rangle,
\end{equation*}
is equal to

\begin{eqnarray}
\label{Prob-1}
    P(\tau_0)=p_0+\sum_{n=1}^{\infty}p_n
    \overline{\vartheta_n},
\end{eqnarray}
where

\begin{eqnarray}
\label{vartheta}
    \vartheta_n\equiv\vartheta_n(\tau_0,\{\delta\})=
    \langle \vartheta_n \big(\tau_0-\tau_n[\{z\},\{\delta\}]\big) \rangle_n.
\end{eqnarray}
Direct calculations similar to those used in the previous subsection (see  Appendix \ref{sec:prob}) lead to the result:

\begin{eqnarray}
\label{vartheta-n5}
    \vartheta_n=
    \theta(\tau_0-\mu_{n,1})\frac{(\mu_{n,1}-
    \tau_0)^n}{\displaystyle{\prod_{k=1}^n} \mu_{n,k}}+
    \frac{W_{n}^{g}\{\mu_{n}\}}{W_{n}\{\mu_{n}\}}, \qquad n\geq 1,
\end{eqnarray}
where

\begin{eqnarray}
\label{vartheta-n6}
    g_j(\tau_0)=\theta(\tau_0-(\mu_{n,1}-\mu_{n,j}))
    \frac{(\tau_0-(\mu_{n,1}-\mu_{n,j}))^n}{\mu_{n,j}}.
\end{eqnarray}
In Eq.(\ref{vartheta-n5}) we used the same notation for $W_n^g$ as in Eqs. (\ref{P-part-7}) - (\ref{P-part-8}). The  scaled transmission time delay $\tilde \tau$ enters into Eq.(\ref{vartheta-n5}) both explicitly and through the functions $g$, Eq.(\ref{vartheta-n6}), while the realization of scatterer intensities $\{\delta\}$ enters by means of variables $\mu$, which, in turn, are determined via the soliton velocities $v_k$ obeying the velocity mapping, Eq. (\ref{VelTransf}).

Another way of obtaining this probability is the direct integration of the probability density function:

\begin{eqnarray*}
\label{vartheta-3}
    \vartheta_n(\tau_0)=\int_0^{\tau_0}\rho_n(\tau) d\tau.
\end{eqnarray*}
The calculation of this integral is reduced to the (straightforward)
integration of the function $f_j$. The latter is equal to:

\begin{eqnarray*}
\label{f-3}
    \int_{0}^{\tau_0}
    f_j (\tau;n) d \tau=
    \theta(\tau_0-\mu_{n,1})\mu_{n,j}^{n-1}
    +\theta(\mu_{n,1}-\tau_0)g_j(\tau_0),
\end{eqnarray*}
and results in the expression:

\begin{eqnarray}
\label{vartheta-4}
    \vartheta_{n}(\tau_0)=
    \theta(\tau_0-\mu_{n,1})+ \theta(\mu_{n,1}-\tau_0)
    \frac{W_{n}^{g}\{\mu_{n}\}}{W_{n}\{\mu_{n}\}}=
    \theta(\tau_0-\mu_{n,1})\left(
    1-\frac{W_{n}^{g}\{\mu_{n}\}}{W_{n}\{\mu_{n}\}}
    \right) +\frac{W_{n}^{g}\{\mu_{n}\}}{W_{n}\{\mu_{n}\}},
\end{eqnarray}
which at first sight looks different from Eq. (\ref{vartheta-n5}) obtained above.
However simple calculations (see Appendix \ref{sec:equiv}) confirm the equivalence of these two formulas.
\section{Conclusion}\label{sec:conc} \thispagestyle{headings}
In this paper we have studied the propagation of the envelope NLSE soliton through a segment containing weak point-like scatterers. Both the positions of scatterers and their intensities were assumed to be random and the concentration of the scatterers was assumed to be small (the mean distance between the scatterers is much larger than the soliton width).

For a relatively short segment, the transmission time delay of a fast soliton is mostly determined by the shifts of the soliton center after each act of scattering. However for sufficiently long segments the main contribution to the delay stems from the shifts of the amplitude and velocity after each scatterer. The crossover lengths separating relatively short segment from a sufficiently long one have also been obtained for both cases of light and heavy solitons.

We have developed a method for calculating the statistical properties of the transmission delay time. This method is applicable not only to the particular problem of the NLSE soliton transmission but also to the problem of forward scattering of an arbitrary classical particle provided that the velocity change during an individual act of scattering is known analytically.

The exact probability density function of the soliton transmission time delay and its two first moments have been found. In the case of identical scatterers we have obtained the PDF as a universal function that depends on a sole parameter - mean number of scatterers in a segment. Thus the first term of the series contains a delta-function singularity, the second has a jump at some point, and each next term contains a jump in the corresponding higher derivative (first, second etc.). The relative fluctuations of the delay time become negligibly small when the segment length grows.

\acknowledgments
\label{sec:acknow}
\thispagestyle{headings}
We are thankful to G.R. Belitski for helpful discussions and M.M. Bogdan for the important criticism. This work was partially supported by a joint scientific project
No.24-02-a 0f NAS of Ukraine and RFBR, Israel Science Foundation
(Grant \# 944/05), the joint French-Ukrainian
project in the framework of scientific cooperation between NASU and the
CNRS. SD, JP and AK would also like to acknowledge the support from the UK Royal Society. SG would like to acknowledge the support obtained from School of Engineering and Applied Science at Aston University, Birmingham, UK for his visit to UK.

\appendix
\section
{The number and energy of waves emitted by a light soliton}
\label{sec:numenergy}\thispagestyle{headings}
In the second order of the perturbation theory with respect to $\ve$, the number $N$ of emitted waves and their energy $E$ are proportional to the following integrals (see Refs. \cite{kkc87,KGSV} and also \cite{BronskiAn}).

\begin{equation}
\label{NewIntegrals-1}
N=\frac{\pi}{2^6}\frac{\ve^2}{v}\int_{-\infty}^{\infty}
F(y,\alpha)dy, \qquad
E=\frac{\pi}{2^8}\ve^2 v
\int_{-\infty}^{\infty}y^2 F(y,\alpha)dy, \ \ \ \alpha=\frac{2a}{v},
\end{equation}
where $F(y,\alpha)$ is the renormalized spectral density of emitted waves:

\begin{equation}
\label{Function-F}
F(y,\alpha)=\frac{\left[ (y-1)^2+\alpha^2\right]^2 }
{\cosh^2 \left[\begin{displaystyle}
\frac{\pi\left(y^2-1+\alpha^2\right)}{4\alpha}
\end{displaystyle} \right]
}.
\end{equation}

Let us sketch a scheme of a general asymptotic analysis of the moment integrals of type (\ref{NewIntegrals-1}),

\begin{equation*}
\label{NewIntegrals-5}
    I_n=\int_{-\infty}^{\infty}y^n F(y,\alpha)dy,
\end{equation*}
for a light soliton $\alpha\ll 1$. In this case the main (power law) contribution comes from the two peaks of the spectral density (\ref{Function-F}), $y_{\pm}=\pm 1+\textit{O}(\alpha^2)$, but the contribution from the right peak, $y_+\simeq 1$, is generally smaller by a factor of $\alpha^4$. In the vicinity of each peak, $y_{\pm}$, one can make a substitution: $$x=\frac{\pi}{4}(y^2+\alpha^2-1),$$ which is of course defined together with its inverse $y_{\pm}(x)$. After the substitution one can write:

\begin{equation*}
\label{NewIntegrals-6}
  I_n=I_n^{+}+I_n^{-}, \qquad  I_n^\pm=\int_{-\infty}^{\infty} f_{\pm}(x,\alpha)\sech^2 \left(x/\alpha\right) \, dx,
\end{equation*}
where

\begin{equation*}
\label{PlusMinus}
    f_{\pm}(x,\alpha)=y_{\pm}^n(x)\left[
    (y_{\pm}(x)-1)^2+\alpha^2\right]^2|y_{\pm}'(x)|.
\end{equation*}
Because the $\alpha$-dependence of the functions $f_{\pm}(x,\alpha)$ is weak (algebraic) while the hyperbolic function is exponentially localized ($\alpha\ll 1$), the asymptotic expansion of the above integral can be performed using an analogue of Watson lemma, i.e. by means of substituting the formal Taylor expansion of $f_{\pm}(x,\alpha)$ at $x=0$ and consequently integrating term by term. The result is then becomes as follows:

\begin{equation*}
\label{NewIntegrals-7}
    I_n^\pm \sim\sum_{k=0}^{\infty}\frac{f_\pm^{(2k)}(0,\alpha)}{(2k)!}
    \int_{-\infty}^{\infty} x^{2k}\sech^2\frac{x}{\alpha}dx=
    \sum_{k=0}^{\infty}\frac{f_\pm^{(2k)}(0,\alpha)}{(2k)!}
    c_{2k}\alpha^{2k+1},
\end{equation*}
where $c_0=2$ and the coefficients $c_{2k}, \ \ k=1,2,...,$ are expressed via Bernoulli numbers $B_{2k}$:

\begin{equation*}
\label{c-k}
    c_{2k}=\frac{2^{2k-1}-1}{2^{2k-2}}\, \pi^{2k} \, \left|B_{2k}\right|.
\end{equation*}

Substituting the corresponding even derivatives of functions $f_\pm$ evaluated at $x=0$ and developing them in series in powers of $\alpha$ we obtain the sought asymptotic expansions for each moment $n$ of the spectrum. The results for $n=0, 2$ yield formulas (\ref{NewIntegrals-4}).
\section{The calculation of moments}\label{sec:Mom}
Consider first a fixed configuration of the scatterer intensities. The functions $f_j(\tau)$, Eq.(\ref{f}), introduced in subsection \ref{subsec:pdf}, divided by $\mu_{n,j}^{n-1}$ become the probability densities themselves. The
two first moments of $\tau$ calculated with these probability densities are:

\begin{eqnarray*}
\label{k}
    \frac{1}{\mu_{n,j}^{n-1}}\int_0^\infty \tau f_j (\tau;n)d\tau =
    \mu_{n,1}-\frac{\mu_{n,j}}{n+1},\ \ \ \ \ \
    \frac{1}{\mu_{n,j}^{n-1}}\int_0^\infty \tau^2f_j
    (\tau;n)d\tau &=&\mu_{n,1}^2-\frac{2\mu_{n,1}\mu_{n,j}}{n+1}
    +\frac{2\mu_{n,j}^2}{(n+1)(n+2)}.
\end{eqnarray*}
Canonical averages $\tau_n^{(1)},$ and $\tau_n^{(2)},$
calculated with the help of canonical probability density
$\rho(\tau,n)$ (see Eq. (\ref{P-part-7}) or  Eq. (\ref{P-part-8})) are:

\begin{eqnarray}
\label{can-means}
    \tau_n^{(1)}=\mu_{n,1}-\frac{
    W_{n}^{(1)}\{\mu_{n}\}
    }{n+1},\ \ \ \
    \tau_n^{(2)}=\mu_{n,1}^2-\frac{2\mu_{n,1}W_{n}^{(1)}
    \{\mu_{n}\}}{n+1}+
    \frac{2 W_{n}^{(2)}\{\mu_{n}\}}{(n+1)(n+2)},
\end{eqnarray}
where $W_{n}^{(m)}\{\mu_{n}\}$ is obtained from
$W_{n}\{\mu_{n}\}\equiv W_{n}^{(0)}\{\mu_{n}\},$ Eq.(\ref{vdM-1}),
by replacing all $\mu_{n,k}^{n-1}$ in the upper row with
$\mu_{n,k}^{n+m-1}.$

All the generalized Vandermonde determinants $W_{n}^{(m)}\{\mu_{n}\}$ are proportional to the initial one $W_{n}\{\mu_{n}\},$ Eq.(\ref{vdM-1}). Corresponding coefficients are expressed as
symmetric polynomials of $\mu_{n,k}$:

\begin{eqnarray*}
\label{symmetr}
    W_{n}^{(1)}\{\mu_{n}\}=W_{n}\{\mu_{n}\}\sum_{k=1}^n
    \mu_{n,k},\ \ \ \
    W_{n}^{(2)}\{\mu_{n}\}=\frac{1}{2}W_{n}\{\mu_{n}\}
    \left[ \left(\sum_{k=1}^n \mu_{n,k}\right)^2-\sum_{1\leq j<k\leq
    n}^n \mu_{n,j}\mu_{n,k}
    \right].
\end{eqnarray*}
Substituting these expressions into Eqs. (\ref{can-means}) after
some straightforward algebra one arrives exactly to Eqs. (\ref{means}) obtained in the body of the paper.
The additional averaging over $\{\delta\}$ completes the calculation.
\section{The cumulative probability distribution}\label{sec:prob}
Explicit form of the  partial probability $\vartheta_n$, see Eq.(\ref{vartheta}),  is:

\begin{eqnarray*}
\label{vartheta-0}
    \vartheta_n=n!\int_0^{\infty}\ldots\int_0^{\infty}D_n(\{z\})
    \theta\left(1-\sum_{k=1}^{\infty}z_k\right)\theta\left(\tau_0-\mu_{n,1}+
    \sum_{k=1}^n\mu_{n,k}z_k\right).
\end{eqnarray*}
With the help of the integral representation of $\theta$-function
it can be written as:

\begin{eqnarray*}
\label{vartheta-n}
    \vartheta_n=\frac{n!}{(2\pi i)^2}
    \int_{C_1}\frac{d\kappa_1}{\kappa_1}e^{i\kappa_1}
    \int_{C_2}\frac{d\kappa_2}{\kappa_2}
    e^{i\kappa_2(\tau_0-\mu_{n,1})}
    \prod_{k=1}^n\int_0^{\infty}e^{iz_k(\mu_{n,k}\kappa_2-\kappa_1)}dz_k.
\end{eqnarray*}
To ensure convergence of all the integrals over $z_k,$ we choose
contours $C_i$ (i=1,2) so that not only $\mathrm{Im}\kappa_i<0$ but also
$\mathrm{Im}(\kappa_1-\mu_{n,k}\kappa_2)<0$ for all $k\leq n$. Then
integration over all $z_k$ leads to the expression
\begin{eqnarray*}
\label{vartheta-n1}
    \vartheta_n&=&\frac{(-1)^n n!}{(2\pi i)^2}
    \int_{C_2}\frac{d\kappa_2}{\kappa_2}
    e^{i\kappa_2(\tau_0-\mu_{n,1})}
    \int_{C_1}\frac{d\kappa_1}{\kappa_1}
    \frac{e^{i\kappa_1}}{\displaystyle{\prod_{k\neq j}^n}
    (\kappa_1-\mu_{n,k}\kappa_2)}.
\end{eqnarray*}
Closing contour of integration of the first the internal integrals
through the upper half plane $\kappa_1$ and taking into account that all
the poles, $\kappa_1=0,\mu_{n,1}\kappa_2,\ldots,\mu_{n,n}\kappa_2,$,
lie within the closed contour $\widetilde{C_1}$, we obtain:

\begin{eqnarray*}
\label{internal}
    \oint_{\widetilde{C_1}}\frac{d\kappa_1}{\kappa_1}
    \frac{e^{i\kappa_1}}{\displaystyle{\prod_{k\neq j}^n}
    (\kappa_1-\mu_{n,k}\kappa_2)}=
    \frac{2\pi i}{\kappa_2^n}\left(
    \sum_{j=1}^n
    \frac{1}{\mu_{n,j}}
    \prod_{k\neq j}^n
    \frac{e^{i\kappa_2\mu_{n,j}}}{\mu_{n,j}-\mu_{n,k}}+
    \frac{(-1)^n}{\displaystyle{\prod_{k=1}^n}\mu_{n,k}}\right).
\end{eqnarray*}
Here the prime over the first product in brackets means that in
the pole (where $k=j$) the second term in the
denominator, $\mu_{n,k}$, must be omitted. As results for $\vartheta_n$ we obtain:

\begin{eqnarray*}
\label{vartheta-n3}
    \vartheta_n=\frac{i^n n!}{2\pi }
    \int_{C_2}\frac{d\kappa_2}{\kappa_2^{n+1}}
    \left\{
    \frac{e^{i\kappa_2(\tau_0-\mu_{n,1})}}{\displaystyle{\prod_{k=1}^n}
\mu_{n,k}}+
    (-1)^n\sum_{j=1}^n \frac{1}{\mu_{n,j}}\prod_{k\neq j}^n
    \frac{e^{i\kappa_2(\tau_0-\mu_{n,1}+\mu_{n,j})}}{\mu_{n,j}-\mu_{n,k}}
    \right\}.
\end{eqnarray*}
Now if we close the contour of integration in the upper plane
$\kappa_2$ and take into account the sole pole of the $(n+1)$-th
order at the origin we arrive at:

\begin{eqnarray}
\label{vartheta-n4}
    \vartheta_n=
    \theta(\tau_0-\mu_{n,1})\frac{(\mu_{n,1}-
    \tau_0)^n}{\displaystyle{\prod_{k=1}^n} \mu_{n,k}}+
    \sum_{j=1}^n
    \frac{
    \theta(\tau_0-(\mu_{n,1}-\mu_{n,j}))(\tau_0-(\mu_{n,1}-\mu_{n,j}))^n
    }{\mu_{n,j}
    \displaystyle{\prod_{k\neq j}^n}(\mu_{n,j}-\mu_{n,k})},
\end{eqnarray}
which is equivalent to Eqs. (\ref{vartheta-n5}), (\ref{vartheta-n6}) presented in Subsection \ref{subsec:mandp}.

\section{Equivalence of the results given by Eqs. (\ref{vartheta-n5}) and (\ref{vartheta-4})}\label{sec:equiv}
To establish the equivalence of these formulas we firstly note that
the last term in the r.h.s. of Eq. (\ref{vartheta-4}) coincides with that of Eq.(\ref{vartheta-n5}). Then the second multiplier in the first term of Eq. (\ref{vartheta-4}) can be presented as:

\begin{eqnarray}
\label{brackets}
    1-\frac{W_{n}^{g}\{\mu_{n}\}}{W_{n}\{\mu_{n}\}}=
    \frac{W_{n}^{h}\{\mu_{n}\}}{W_{n}\{\mu_{n}\}},
\end{eqnarray}
where

\begin{eqnarray*}
\label{h}
    h_j(\tau_0)=\mu_{n,j}^{n-1}-
    \frac{(\tau_0-(\mu_{n,1}-\mu_{n,j}))^n}{\mu_{n,j}}.
\end{eqnarray*}
The last ratio in Eq.(\ref{brackets}) is a $n$-th power polynomial
of $\tau_0$ that vanishes together with all its first $n-1$
derivatives at the point $\tau_0=\mu_{n,1}$. Therefore one gets:

\begin{eqnarray*}
\label{P}
    \frac{W_{n}^{h}\{\mu_{n}\}}{W_{n}\{\mu_{n}\}}=
    C(\tau_0-\mu_{n,1})^n.
\end{eqnarray*}
To find the constant $C$, we differentiate this identity $n$ times
and obtain:

\begin{eqnarray*}
\label{C}
    C=-\sum_{j=1}^{n}\frac{\displaystyle{\frac{1}{\mu_{n,j}}}}
    {\displaystyle{\prod_{k\neq j}}
    (\mu_{n,j}-\mu_{n,1})}
    =-\frac{W_{n}^{\frac{1}{\mu_{n,j}}}\{\mu_{n}\}}{W_{n}\{\mu_{n}\}}.
\end{eqnarray*}
Recall that $W_{n}^{\frac{1}{\mu_{n,j}}}$ is defined by Eq. (\ref{vdM-2}). The numerator in the last ratio can be easily calculated and equals to

\begin{eqnarray*}
\label{W}
    W_{n}^{\frac{1}{\mu_{n,j}}}\{\mu_{n}\}=(-1)^{n-1}
    \frac{W_{n}\{\mu_{n}\}}
    {\displaystyle{\prod_{k=1}^{n}}
    \mu_{n,k}}.
\end{eqnarray*}
This result together with two previous equations establishes the
identity of the two forms of probability $\vartheta_n(\tau_0)$ given by
Eqs.(\ref{vartheta-n5}) and (\ref{vartheta-4}).


\begin{thebibliography}{99}
\bibitem{NL&D}A.R. Bishop, D.K. Campbell, S. Pnevmatikos (Eds.),
\textit{Disorder and Nonlinearity}, Springer, Berlin, 1989;\\
    F. Abdullaev, A.R. Bishop, S. Pnevmatikos (Eds.), \textit{Nonlinearity with Disorder}, Springer, Berlin, 1992;\\
    A.R. Bishop, S.Jimenez, L. Vazquez (Eds.), \textit{Fluctiation Phenomena: Disorder and Nonlinearity}, Word Scientific, Singapore, 1995.

\bibitem{KG92} S. A. Gredeskul and Yu. S. Kivshar, Phys. Rep. \textbf{216}, 1 (1992).

\bibitem{HaifaWorkshop}
    T. Schwartz, G. Bartal, S. Fishman, M. Segev,
    Nature (London) \textbf{446}, 52 (2007);\\
    A. Iomin, S. Fishman,
    Phys. Rev. E \textbf{76}, 056607 (2007);\\
    Y. Lahini, A. Avidan, F. Pozzi, M. Sorel, R.  Morandotti,
D. N. Christodoulides, Y. Silberberg,
    Phys. Rev. Lett. \textbf{100}, 013906 (2008);\\
    G. Kopidakis, S. Komineas,S. Flach, S. Aubry,
    Phys. Rev. Lett. \textbf{100}, 084103 (2008);\\
    A.S. Pikovsky, D.L. Shepelyansky,
    Phys. Rev. Lett. \textbf{100}, 094101 (2008).

\bibitem{KGSV} Yu.S. Kivshar, S.A. Gredeskul, A. Sanchez, L. Vasquez,
    Phys. Rev. Lett. \textbf{64}, 1693 (1990).

\bibitem{Hopkins} V.A. Hopkins, J. Keat, G.D. Meegan, T. Zhang,
J.D. Maynard, Phys. Rev. Lett. \textbf{76}, 1102 (1996).

\bibitem{Devillard} P. Devillard, B. Souillard,
    J. Stat. Phys. \textbf{43}, 423 (1986);\\
    B. Doucot, R. Rammal,
    J. Physique \textbf{48}, 509 (1987);\\
    R. Knapp, G.P. Papanicolaou, B. White,
    J. Stat. Phys. \textbf{63}, 567 (1991).

\bibitem{zs71}
    V.E. Zakharov, A.B. Shabat, Zh. Exp. Teor. Fiz.
    \textbf{61}, 118 (1971)
    [Sov. Phys. JETP \textbf{34}, 62 (1971)].

\bibitem{mnp84} S.V. Manakov, S.P. Novikov,
    L.P. Pitaevskii, and V.E. Zakharov,
    \textit{Theory of Solitons},
    Consultants Bureau, New York, 1984).

\bibitem{km89} Y.S. Kivshar, B.A. Malomed,
    Rev. Mod. Phys. \textbf{61}, 763 (1989).

\bibitem{kkc87}
    Yu.S. Kivshar, A.M. Kosevich, O.A. Chubykalo,
    Fiz. Nizkih Temp. \textbf{13}, 438 (1987)
    [Sov. J. Low Temp.Phys. \textbf{13}, 251 (1987)];\\
    A.M. Kosevich, Yu.S. Kivshar, O.A. Chubykalo,
    in \textit{Lecture Notes in Physics:
    Physics of Phonons}, Ed. by T. Paszkievich,
    Springer-Verlag, vol.285, p.419(1987).

\bibitem{Burtsev}
    S. Burtsev, D.J. Kaup, B.A. Malomed,
    Phys. Rev. E \textbf{52}, 4474 (1995).

\bibitem{Knapp}
    R.Knapp, Physica D \textbf{85}, 496 (1995).

\bibitem{BronskiAn}
    J.C. Bronski, J. Nonlinear Sci. \textbf{8}, 161 (1998).

\bibitem{BronskiNum}
    J.C. Bronski, J. Stat. Phys. \textbf{92}, 995 (1998).

\bibitem{GarnierSIAM} J. Garnier,
    SIAM J. Appl. Math. \textbf{58}, 1969 (1998).

\bibitem{GarnierWRM} J. Garnier,
    Waves Random Media \textbf{11}, 149 (2001).

\bibitem{KFV} Yu.S. Kivshar, Z. Fei, L. Vasquez,
    Phys. Rev. Lett. \textbf{67}, 1177 (1991).

\bibitem{FKV0} Z. Fei, Yu.S. Kivshar, L. Vasquez,
    Phys. Rev. A, \textbf{45}, 6019 (1992).

\bibitem{Gonz03} J.A. Gonzalez, A. Bellorin, L.E. Guerrero,
    Chaos, Solitons and Fractals \textbf{17}, 907 (2003).


\bibitem{FKV1} Z. Fei, Yu.S. Kivshar, L. Vasquez,
    Phys. Rev. A, \textbf{46}, 5214 (1992).

\bibitem{Kalbermann} G. K\"albermann,
    Phys. Rev. E \textbf{55}, R6360 (1997).



\bibitem{kik83} A.M. Kosevich, B.A. Ivanov, A.S. Kovalev,
    \textit{Nonlinear Magnetization Waves.
    Dynamical and Topological Solitons},
    Naukova Dumka, Kiev (1983);
    Phys. Rep. \textbf{194}, 117 (1990).

\bibitem{KA} Yu.S. Kivshar and G.P. Agrawal,
      \textit{Optical Solitons: From Fibers to Photonic Crystals},
      Academic Press, San Diego   (2003).

\bibitem{dty05} S.A. Derevyanko, S.K. Turitsyn and D.A. Yakushev, J. Opt. Soc. Am. B \textbf{22}, 743 (2005).

\bibitem{fiberarrays} Y.V. Kartashov, V.A. Vysloukh, and L. Torner, Opt. Lett. \textbf{33}, 1747 (2008);\\ Y.V. Kartashov, V.A. Vysloukh, and L. Torner, \pra \textbf{77}, 051802 (2008).

\bibitem{KPGD} A.S. Kovalev, J.E. Prilepsky, S.A. Gredeskul,
S.A. Derevyanko, Fiz. Nizk. Temp. \textbf{34},707 (2008) [Low. Temp. Phys. \textbf{34}, 559 (2008)].

\bibitem{km77} V.I. Karpman, E.M. Maslov,
    Zh. Eksp. Teor. Fiz. \textbf{73}, 537 (1977)
    [Sov. Phys. JETP \textbf{46}, 281 (1977)];\\
    D.J. Kaup,
    SIAM J. Appl. Math. \textbf{31}, 121 (1976);\\
    D.J. Kaup, A.C. Newell,
    Pro. R. Soc. A \textbf{361}, 413 (1978).
\end{thebibliography}
\end{document}